\documentclass[12pt,preprint]{aastex}

\newcommand{\myemail}{zhekovs@colorado.edu}
\newcommand{\NGC}{NGC~6888~}
\newcommand{\NGCE}{NGC~6888}
\newcommand{\kms}{~km s$^{-1}$~}
\newcommand{\dotM}{~M$_{\odot}$~yr$^{-1}$~}

\shorttitle{X-rays from \NGCE}
\shortauthors{Zhekov and Park}

\begin{document}


\title{
{\it Suzaku} Observations of the 
Prototype Wind-Blown Bubble
\NGC 
}



\author{Svetozar A. Zhekov\altaffilmark{1,3},
and Sangwook Park\altaffilmark{2} }

\altaffiltext{1}{JILA, University of Colorado, Boulder, CO
80309-0440, USA; \myemail}
\altaffiltext{2}{Department of Physics, Box 19059, 
University of Texas at Arlington,  Arlington, TX 76019;
s.park@uta.edu}
\altaffiltext{3}{On leave from Space Research Institute, Sofia,
Bulgaria}


\begin{abstract}
We present an analysis of the {\it Suzaku} observations of the
prototype wind-blown bubble \NGC which is based both on use of
standard spectral models and on a direct comparison of
theoretical models with observations. 
The X-ray spectra of \NGC are
soft and most of the X-rays are in the (0.3 - 1.5 keV) energy range.
But, hard X-rays (1.5 - 4.0 keV) are also detected ($\sim 10$\% of the
observed flux).
The corresponding spectral fits require a relatively cool plasma with
kT~$\leq 0.5$~keV but much hotter plasma with temperature kT~$\geq
2.0$~keV is needed to match the observed hard X-ray emission.
We find no appreciable temperature variations within the hot bubble
in \NGCE. The derived abundances (N, O, Ne) are consistent with those
of the optical nebula.  This indicates a common origin of the
X-ray emitting gas and the outer cold shell:  most of the X-ray
plasma 
(having non-uniform spatial distribution: clumps)
has flown into the hot bubble from the optical nebula.
If the electron thermal conduction is efficient, this can naturally
explain the relatively low plasma temperature of most of the X-ray
emitting plasma. Alternatively, the hot bubble in \NGC will be
adiabatic and the cold clumps are heated up to X-ray temperatures
likely by energy exchange between the heavy particles (hot ions
diffusing into the cold clumps).

\end{abstract}





\section{Introduction}
Optical nebulosities trace large 
cavities around early-type stars (O, Of and Wolf-Rayet, WR) that are 
formed as the high-speed wind sweeps up ambient interstellar gas and 
compresses it into a thin shell. 
The flow pattern, resulting from this interaction, was first recognized
by Pikelner (1968) and consists of two regions of 
shocked gas: one corresponding to the shocked stellar wind and 
the other to shocked interstellar gas. The two regions are separated 
by a contact discontinuity and the hot interior gas can cool via 
thermal conduction across the interface. Because of its high 
temperature, the interior of the wind-blown bubbles (WBB) is expected 
to emit in X-rays.

In 1970’s and 1980’s, analytical solutions for the WBB structure were
derived that revealed details in the physics of these interesting
objects. For a review of the analytical works and the related physical
ideas see Dyson (1981) and McCray (1983). It should be noted that the
most complete (semi-)analytical study of the WBB structure was
presented in  Weaver et al. (1977).
With the increasing computing power, the
numerical hydrodynamic simulations became a powerful tool for studying
the physics of these objects.
The first numerical modeling of WBB was done by \citet{falle_75}
and later on details of the adiabatic, radiative and conductive 
WBB were studied numerically  
(e.g., \citealt{ro_85}; Rozyczka \& Tenorio-Tagle, 1985a,b,c;
Brighenti \& D'Ercole, 1995a,b, 1997; \citealt{de_92}; \citealt{gs_95};
\citealt{gs_96a}; \citealt{gs_96b}; \citealt{zhm_98}, 2000).
Thanks to numerical simulations, it was found that the WBB
are subject to numerous dynamic instabilities. These simulations
also allowed a much more complete and complex physical picture be
explored, namely, by following various phases of the WBB evolution 
when the wind of the central stars varies with time (for details see
\citealt{gs_96a}; \citealt{gs_96b}).

Observational properties of the optical nebula in WBB were described in 
detail in the early works of Chu et al. (1983 and the references 
therein) and Lozinskaya (1982; see also the book by Lozinskaya 1993 
for an observational review).
But, it should be emphasized that the physics of the hot interior 
(hot bubble) of a wind-blown bubble is a cornerstone in the entire
physical picture of these objects. This is so since the hot bubble is 
the region where the stellar wind energy is stored and subsequently 
used for driving the entire structure. From such a point of view, X-ray
observations of WBB are very important because they can provide us
with details about the physical conditions in the hot bubble.

The first successful X-ray detection of a WBB was that of \NGC by
{\it Einstein} \citep{boch_88}. It was found that the X-ray emission 
from NGC 6888 is characterized by a plasma temperature 
kT $= 0.28 - 0.8$~keV (90\% confidence interval), or T $= 3.2 - 9$~MK.
{\it ROSAT} observations
were sensitive to cooler plasma and yielded a characteristic
temperature T $\approx 2$~MK. They showed that the X-ray emission arises 
primarily in filament like structures \citep{wri_94}.
{\it ASCA} observations suggested that even hotter plasma at 
T $\approx 8$~MK might be present in the bubble interior 
\citep{wri_05}. Preliminary results from a recent Chandra 
observation, {\it which covered only the northeast part} of \NGCE, 
detected plasma with a temperature of T $\approx 2-3$~MK and possible 
nitrogen enrichment \citep{chu_06}.
Similar soft X-ray spectra were detected for another WBB, S308, with 
{\it ROSAT} \citep{wri_99} and {\it XMM-Newton} \citep{chu_03}. 
Unfortunately, none of these data provided us with solid grounds to
draw a firm conclusion about the physical mechanism
responsible for the X-ray emission from WBB. Obviously, 
the X-ray emitting plasma in the hot bubble has a moderate temperature
(considerably lower than the one expected from a shock with 
velocity equal to that of the stellar wind)
and the reason for this could be that the thermal conduction
operates efficiently in the hot interior. 
But, it is necessary to have solid observational arguments that
support or rule out such a theoretical expectation. 

This was the basic motivation for the recent {\it Suzaku} observations 
of \NGC which covered the entire object and provided us with the X-ray 
spectra having good photon statistics. We report here the results
from our analysis of these data. The paper is organized as follows.
We give some basic information about the WBB \NGC in
Section \ref{sec:thebubble}. In Section \ref{sec:observations},
we briefly review the {\it Suzaku} observations. In Section
\ref{sec:global}, we present results from the global spectral
models. In Section \ref{sec:origin}, we discuss the origin of the
hot gas in \NGCE. In Section \ref{sec:discussion}, we discuss our
results and we present our conclusions in Section
\ref{sec:conclusions}.

\section{The Wind-Blown Bubble \NGC}
\label{sec:thebubble}
The nebula \NGC around the Wolf-Rayet star WR 136 is the prototype 
of a broad class of ring nebulae around stars with strong stellar winds. 
The size of the optical nebula is $12' \times 18'$ and its expansion 
velocity is 75\kms \citep{chu_83}.
The distance to \NGC (WR 136) is d~$= 1.26$~kpc \citep{vdh_01}, thus,
the dynamical age of the object is 30,000 years.
The observed X-ray flux in the {\it ROSAT} PSPC band is 
$1.2\times10^{-12}$ erg cm$^{-2}$ s$^{-1}$ 
with an X-ray absorption column density of 
$N_H = 3.2 \times 10^{21}$ cm$^{-2}$ \citep{wri_94}. 
A larger flux value of $1.86\times10^{-12}$ erg cm$^{-2}$ s$^{-1}$ is 
obtained from  the {\it ACSA} observations of this object
\citep{wri_05}.

The central star, WR 136, is a WN6 star \citep{vdh_01} with stellar
wind parameters V$_{wind} = 1600$\kms \citep{prinja_90} and 
$\dot{M} = 1.5\times10^{-5}$\dotM. For the mass-loss rate, we used
the result for WR 136 derived by \citet{ab_86}, $\log \dot{M} =
-4.5$~\dotM for d~$= 1.8$~kpc and V$_{wind} = 2000$\kms,
and corrected it for
the values of the wind velocity and distance adopted here. We made
use of the general dependence $\dot{M} \propto V_{wind}\, d^{1.5}$
for the mass-loss values derived from the radio
observations (\citealt{pf_75}; \citealt{wb_75}).

\section{Observations and Data Reduction}
\label{sec:observations}
\NGC was observed with {\it Suzaku} in two runs: June 4 and November 
16, 2009, centered on the North-East and South-West parts of this 
object thus covering the entire optical nebula. The basic data for our
analysis are those from the three active X-ray Imaging Spectrometers 
(XIS) on-board {\it Suzaku}. We note that one of them (XIS1) is
back-illuminated while the other two (XIS0, XIS3) are 
front-illuminated CCDs which allows the extracted spectra from units 
XIS0 and XIS3 be combined for further analysis
\footnote{\label{foot:1} see The 
{\it Suzaku} Data Reduction Guide: 
http://heasarc.gsfc.nasa.gov/docs/suzaku/analysis/abc/
}.
We extracted X-ray spectra of \NGC from the filtered and screened
pipeline events files using the $3\times3$ and $5\times5$
observational modes
\footnote{\label{foot:2} In the $3\times3$ and $5\times5$ editing modes,
all the pulse heights of the corresponding 9 and 25 pixels centered 
at the event center are sent to the telemetry. The data for each mode
come from different times which effectively increases the exposure
time. See
http://heasarc.gsfc.nasa.gov/docs/astroe/prop\_tools/suzaku\_td
}
 which resulted in 
88.8 ksec and 77.3 ksec effective exposures for the June and November
observations, respectively.
The total number of source counts in the (0.3 - 4 keV) energy band are 
5403 (XIS1) and 4838 (XIS0$+$XIS3) for the North-East half;
4330 (XIS1) and 3413 (XIS0$+$XIS3) for the South-West half of
\NGCE.
For the current analysis, we extracted one total and  two subregion 
spectra for each of the observed parts of \NGCE.
The corresponding X-ray images as well as source and background
regions for the spectral extraction are shown in
Fig.~\ref{fig:images}.
The circles in red denote the regions around bright point sources
(No. 5, 6 and 7 in Table 4 of Wrigge et al. 1994) that were excluded 
from the spectral extraction. We note that source No.5 is in fact the
central star (WR 136) in the nebula.  Since it is not a strong X-ray 
source \citep{sk_10}, the presence of a near by point source (No.6)
and the large PSF of {\it Suzaku} telescopes prevents extraction of
valuable information on its X-ray emission.
Despite the large `pixel' of {\it Suzaku} images,
Figure~\ref{fig:images} shows that the X-rays of \NGC originate from
region {\it inside} the optical nebula and their spatial distribution
is {\it not} uniform (e.g., indicating clump like structures).

We adopted 
the current calibration database for {\it Suzaku}, XIS (20100123) and
XRT (20080709), to generate the corresponding response matrices  and
ancillary  response files.
We made use of standard as
well as custom models in version 11.3.2 of XSPEC
\citep{Arnaud96}
for the spectral analysis in this study.

Finally, to facilitate comparisons of the new results on \NGC with 
those from the previous X-ray observations of this object, we mention 
a few basic characteristics of the {\it Suzaku} observations and 
similarly for those of {\it ROSAT} and {\it ASCA}. 
The {\it Suzaku} telescopes have spatial resolution 
of $\sim 2$~arcmin (expressed as Half-Power Diameter, or HPD).
The XIS detectors operate in the $0.2-12$~keV energy range. 
At energies below 1 keV the back-illuminated CCD (XIS1) has much 
better sensitivity (at least a factor of 2) than the front-illuminated 
CCDs (XIS0, 3), while their effective area is about the same at higher
energies (see \S~3 in The {\it Suzaku} Technical Description,
footnote \ref{foot:2}).
On the other hand
\footnote{\label{foot:3} 
For  the {\it ASCA} Technical Description see\\
http://heasarc.gsfc.nasa.gov/docs/asca/asca\_nra06/appendix\_e/appendix\_e.html.
\\
For the {\it ROSAT} Technical Description see
http://heasarc.gsfc.nasa.gov/docs/rosat/ruh/handbook/
and 
http://heasarc.gsfc.nasa.gov/docs/journal/pspc2.html
},
the energy bandpass and spatial resolution (HPD)  were: 
$0.5-12$~keV and $2.9$~arcmin for {\it ASCA}; 
$0.12-2.4$~keV and $\sim 30$~arcsec for {\it ROSAT}.





\section{Global Spectral Models}
\label{sec:global}

\subsection{Soft X-ray Spectra}
\label{subsec:soft}
A general property of the {\it Suzaku} spectra of \NGC 
is that they are rather soft and no apparent line features 
are detected at photon energies $\geq 1.5$~keV. Hence, we
started our analysis with the X-ray spectra in the (0.3 - 1.5 keV) range
(re-binned to have a minimum 100 counts per bin)
where most of the X-ray counts ($\sim 90$\%) are found. We adopted
the following approach. 

We used canonical discrete-temperature plasma models to derive the
global properties of the X-ray emitting plasma in the hot bubble of
\NGCE. 
Since the ionization history of the plasma in the hot bubble is very
complex and not well constrained,
we considered two limiting cases: models that assume plasma in 
collisional ionization equilibrium (CIE) or plasma with non-equilibrium 
ionization (NIE).  For consistency in the atomic data, we made  use
of emission models {\it vapec} and {\it vpshock} in XSPEC.

As a first step, we fitted the total spectra of the North-East 
(combined region 1 and 2 in Fig.~\ref{fig:images}) and South-West 
(combined region 3 and 4 in Fig.~\ref{fig:images}) halves 
simultaneously (four spectra in total). Although the quality of the
{\it Suzaku} spectra is much better compared to that of the previous
studies of the entire object, using {\it ROSAT} and {\it ASCA}
(\citealt{wri_94}; \citealt{wri_05}), we do not find it sufficient to
allow us investigate possible variations of the X-ray absorption or
abundances. Thus, we assumed uniform abundances for the hot plasma in 
\NGC and its X-ray emission is subject to the same interstellar
absorption. These were free parameters in our fits. 
As a basic set of abundances, we adopted the solar abundances of 
\citet{an_89}. We varied only the  N, O, Ne, Mg and Fe abundances in
the spectral fits since these are the elements that provide most of 
the X-rays in the (0.3 - 1.5 keV) energy range. 
We also explored a different set of (non-solar) abundances
(\citealt{vdh_86}; see below) that is typical for the stellar wind of 
the central WN6 star, since this star is the energy source for the hot 
bubble in \NGCE.

\begin{figure}[ht]
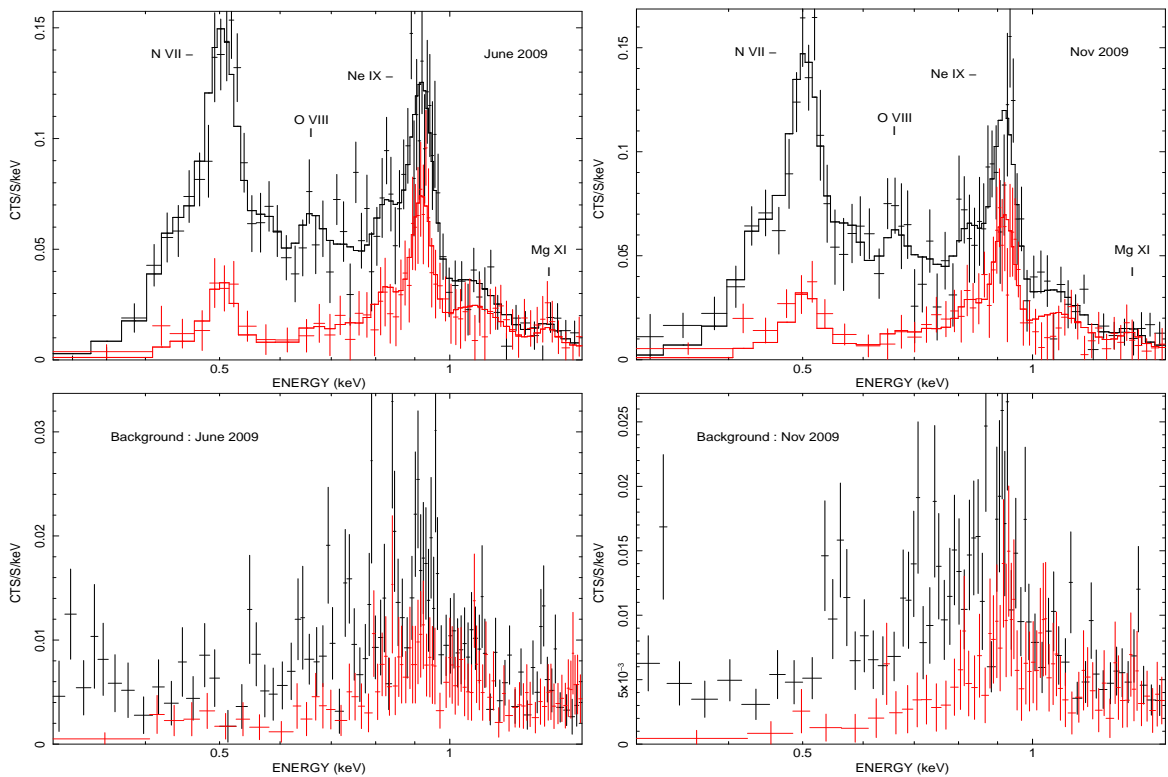

\centering\includegraphics[width=2.in,height=3.in,angle=-90]{f2a.eps}
\centering\includegraphics[width=2.in,height=3.in,angle=-90]{f2b.eps}
\centering\includegraphics[width=2.in,height=3.in,angle=-90]{f2c.eps}
\centering\includegraphics[width=2.in,height=3.in,angle=-90]{f2d.eps}
\caption{
\NGC background-subtracted spectra overlaid with the best-fit 
1T shock model (Table~\ref{tab:fits}). In each upper panel, the XIS1
spectrum is in black and the XIS0$+$3 spectrum is in red (lower
curve). The North-East and South-West halves of the bubble were
observed correspondingly in June and November 2009.
For comparison, the lower panels show the corresponding background spectra 
after subtracting the non X-ray background. 
}
\label{fig:spectra}
\end{figure}

The results from the global model fits to the entire X-ray spectrum of
\NGCE, assuming uniform plasma temperature, are given in
Table~\ref{tab:fits} and Fig.~\ref{fig:spectra}.
We note the good quality of the fits and the fact that these fits
confirm the early finding that the X-ray plasma in \NGC is relatively
`cool' (\citealt{boch_88}; \citealt{wri_94}; \citealt{wri_05};
\citealt{chu_06}). The value of the X-ray absorption is slightly
smaller than the one found from the analysis of the entire X-ray
spectrum using {\it ROSAT} and {\it ASCA} (\citealt{wri_94};
\citealt{wri_05}) while the observed flux in the {\it Suazku} spectra
of \NGC is by 10-20\% higher. A possible reason for this could be the
quality of the spectra and we note that the total number of 
counts of the {\it Suzaku} spectra was at least one order of magnitude 
larger than that in the {\it ROSAT} and {\it ASCA} spectra 
of \NGCE. Thus, we believe that the results from the current analysis
are better constrained.
For comparison, Figure~\ref{fig:spectra} also shows the background
spectra after subtracting the non X-ray background, NXB (the NXB
spectra were constructed following the procedure in  the {\it Suzaku} 
Data Reduction Guide, see footnote \ref{foot:1}  in 
\S~\ref{sec:observations}).
We see that the source and background spectra are very different
(e.g., no N VII emission in the latter).
This makes us confident about the derived results,
despite the large `pixel' of the {\it Suzaku} telescopes and the close 
location of the source and background subtraction regions.

As a next step, we searched for temperature variations of the X-ray
plasma in the hot bubble of \NGCE. For this goal, we used 1T {\it
vpshock} model to fit the
individual spectra of region 1 and 2 in the North-East and of 
region 3 and 4 in the South-West halves (see Fig.~\ref{fig:images})
simultaneously (eight spectra in total). All the spectra shared the
same abundances and X-ray absorption. We note that the derived
X-ray absorption (N$_H = 2.40^{+0.50}_{-0.38}\times 10^{21}$
cm$^{-2}$) and 
abundances are consistent (being within $1\sigma$ errors) with their 
values from the spectral model fits with uniform plasma temperature 
and there is no appreciable variation in the plasma temperature
within the hot bubble: kT$= 0.47^{+0.05}_{-0.12}$ (region 1);
$0.43^{+0.13}_{-0.10}$ (region 2); 
$0.43^{+0.11}_{-0.09}$ (region 3);
$0.37^{+0.05}_{-0.09}$ (region 4);
that is all the values are within the $1\sigma$ errors from each 
other. The same is valid for the observed fluxes in the 
(0.3 - 1.5 keV) energy range: F$_X = (4.44; 4.69; 4.79
\mbox{~and~} 4.70) \times 10^{-13} $ ergs cm$^{-2}$ s$^{-1}$ for
region 1, 2, 3 and 4, respectively.
The observed spectra overlaid with the best-fit 1T shock model
($\chi^2/dof = 196/232$) are
shown in Fig.~\ref{fig:spectra_quart}.
Although the X-ray emission of \NGC originates
mostly from a non-uniform distribution of hot plasma (filaments,
clumps; see 
\citealt{wri_94}; \citealt{wri_05}; see also Fig.~\ref{fig:images}), 
the properties of the X-ray emitting gas (temperature, flux) are quite 
uniform through the entire hot bubble.
So, the high signal-to-noise {\it Suzaku} data confirm similar 
findings already reported from the analysis of the {\it ROSAT} and 
{\it ASCA} observations of \NGC (\citealt{wri_94}; \citealt{wri_05}).
But, we caution that the almost equal values for the 
observed flux from regions 1, 2, 3 and 4 do not mean that 
the plasma emissivity in the hot bubble is uniform. As seen from
Fig.~\ref{fig:images}, regions 1 and 4 comprise much smaller part 
from the nebula than regions 2 and 3 do. Therefore, the plasma
emissivity (or emission measure) of the X-ray emitting plasma is
higher in regions 1 and 4, likely due to higher plasma density
or higher concentration of clumps and filaments.

It is interesting to note that although we used simple plasma models
to fit the observed X-ray spectra of \NGCE, the derived abundances
are fairly consistent between different models, and 
they  are generally {\it inconsistent}
with the typical abundances of a WN star (see below).
In addition to the
two basic models (2T {\it vapec} and {\it vpshock})  used to fit the 
entire spectrum of \NGC, we ran some more complex ones: (i) a model 
with a distribution of thermal CIE plasma (we adopted our custom model
which is similar to {\it c6pvmkl} in XSPEC but uses the {\it apec}
collisional plasma for the X-ray spectrum at a given temperature);
(ii) a model with a distribution of adiabatic shocks with NIE effects
taken into account (we made use of our custom model for XSPEC which
was successfully used in the analysis of the X-ray spectra of SNR~1987A
- e.g., \citealt{zh_09} and the references therein).
We will discuss the result from these more complex models below (see
\S~\ref{subsec:hotplasma}), and we only mention that the models
with a distribution of CIE plasma or NIE shocks show two peaks in 
the distribution of their emission measure with temperature: one for
a `cool' plasma ( kT $< 0.4$ keV) that dominates the X-ray emission
from \NGC and another for a `hot' plasma (kT $\geq 2$ keV).
The values for the plasma temperature of the 2T {\it vapec} model are
consistent with the values of the temperature peaks in the
distribution of the CIE plasma.

For comparison, we also ran a {\it vpshock} model with a set of typical
WN abundances \citep{vdh_86}. We note that there was no good fit to
the observed spectra unless we varied the abundances  of N, O, Ne, Mg,
and Fe. 
Figure~\ref{fig:abund} presents the derived abundances from various
model fits to the entire X-ray spectrum of \NGCE. All abundances are
given with respect to the total number density of hydrogen and helium
(H$+$He) which facilitates the comparison between cases with
completely different basic set of abundances (e.g. solar vs.
non-solar).
On the one hand, this result illustrates the uncertainties of the
derived abundances if using different plasma models and, on the other
hand, it conclusively demonstrates that the chemical composition of the 
X-ray emitting plasma in \NGC is quite different from that of the
stellar wind of the central star in this object.

\subsection{Very Hot Plasma in \NGCE?}
\label{subsec:hotplasma}
In order to search for indications of X-ray emission from hot plasma
(kT $\geq 2$~keV), we expanded the photon range of interest and
considered the X-ray spectra of \NGC in the (0.3 - 4.0 keV) energy 
range. This increased the total number of source photons by $\sim
10$\% compared to the (0.3 - 1.5 keV) spectra analyzed above.
To improve the photon statistics, we re-binned the spectra to have a
minimum of 200 counts per bin. 

We used two models to fit the spectra: (i) one with a distribution of 
CIE plasma; (ii) and a model with a distribution of NEI shocks (as 
described in 
\S~\ref{subsec:soft}). The quality of the fits was very good:
$\chi^2/dof = 122/169$ (CIE plasma) and $\chi^2/dof = 120/168$ (NEI
shocks). The derived values for the X-ray absorption and chemical
abundances were entirely consistent with the values discussed in our
analysis of the soft (0.3 - 1.5 keV) spectra.
Figure~\ref{fig:spectra_hot} shows the \NGC spectra overlaid with the
best-fit model for a distribution of CIE plasma as well as the derived
distribution of emission measure for the CIE plasma and NEI shocks.
%

We see that both models show a bi-modal distribution of the X-ray
emitting plasma with peaks quite distinguished in the temperature
space. The `cool' plasma (kT~$\leq 1$~keV) is the dominant source of
X-ray emission in the (0.3 - 1.5 keV) energy range but provides only
between (6 - 19)\% of the flux at higher energies (1.5 - 4.0 keV),
depending on the model used (e.g., CIE plasma or NIE shocks).
It is thus quite likely that some hot plasma is present in the
X-ray emitting region of \NGCE. This plasma is needed to explain 
the weak but statistically significant emission 
(signal-to-noise~$\sim 6 - 15$) in the (1.5 - 4.0 keV) 
energy range which may have been missed in the {\it ROSAT} and 
{\it ASCA} spectra of the entire WBB due to their 
considerably poorer photon statistics, 
and almost no sensitivity at energies $\geq 2$~keV for {\it ROSAT}.

\section{The Origin of the Hot Gas in \NGC}
\label{sec:origin}
The two most important results from the model fits to the
X-ray spectra of \NGC that could be directly related to the origin of
the hot gas in this object are:  (i) the X-ray
plasma in \NGC has a relatively low temperature; (ii) the hot bubble 
abundances are not consistent with the ones of the stellar wind of the 
central star in \NGC (see \S~\ref{sec:global}).

The latter is justified in two ways as seen from Fig.~\ref{fig:abund}. 
The X-ray abundances are smaller than the corresponding 
canonical WN values (typical for the central star) by (0.5 - 1.5) dex
on the average. And, the N/O ratio derived from the
analysis of the {\it Suzaku} data has a characteristic value of
2.0 - 3.0 which is much lower than the value of  $\sim 21.5$ for
the N/O ratio in the WN abundances set (e.g., \citealt{vdh_86}). 
We note that low values ($\sim 2$) for the N/O ratio were derived 
from analysis of the optical emission of \NGCE. Also, the 
N, O and Ne abundances derived from X-rays are very similar to those 
reported from the optical (\citealt{kwi_81}; \citealt{moore_00}; 
see also Fig.~\ref{fig:abund}).

It is thus natural to conclude that the plasma which dominates the 
X-ray emission from \NGC likely originates from the cold optical nebula.
Subsequently, this cold plasma  has been heated to high temperatures 
through a 
mechanism that is yet to be identified. This heating mechanism must be 
capable of explaining the relatively low plasma temperature in the hot 
bubble of \NGC (kT $< 0.5$~keV; see \S~\ref{sec:global}).

As proposed by the standard WBB model
(\citealt{weaver_77}; for numerical modeling see \citealt{zhm_98}), 
the electron thermal conduction is such a physical mechanism.
Alternatively, the models of so called mass-loaded astronomical flows
also suggest a relatively low temperature in the WBB objects
(e.g., \citealt{hartquist_86}; \citealt{arthur_93}) but we note that
it is not well established which is the actual physical mechanism that
is responsible for the `instantaneous' mixing of the hot bubble gas 
with the mass-loaded flow.  
In either case some important questions must find their 
answers in order to build a more complete global picture of \NGCE.

Namely, what is the fate of the shocked stellar wind? 
Given the age of this WBB (\S~\ref{sec:thebubble}), we note 
that the massive WN wind can supply up to one solar mass or so of hot 
gas with distinct non-solar abundances. How can we explain that
we find no indications for such a plasma in the X-ray spectrum
of \NGCE? 
And, how can the presence of really high-temperature plasma
(kT~$\geq 2$~keV) fit into a general physical picture for WBB?
Below, we will discuss these issues in some extent.






\section{Discussion}  
\label{sec:discussion}

\subsection{Wind-Blown Bubble Hydrodynamics}
\label{subsec:hydro}
In order to have a realistic view on the fate of the shocked WN
stellar wind, we need to carry a direct comparison between the
theoretical X-ray emission and the observed spectra of \NGCE. 
For this goal, we adopted the following procedure. First,
we used a 1D numerical hydrodynamic model to derive the distribution of 
plasma characteristics (temperature, density) in the hot bubble. In
turn, we constructed the hot plasma distribution of emission measure 
with temperature. Second, we developed  a custom model for XSPEC that 
uses thus derived distribution of emission measure to model the X-ray 
emission from \NGCE. Finally, we used this model to fit the {\it
Suzaku} spectra and check the consistency of the physical model
with observations: e.g., whether the model provides correct observed
flux, abundances, X-ray absorption.

We briefly review some details about the WBB hydrodynamics (see
\citealt{weaver_77} and  numerical modeling in \citealt{zhm_98}). 
When a massive stellar wind interacts with the circumstellar matter,
CSM, 
a two-slab 
structure will form, bounded by two shocks and a contact discontinuity 
separating the shocked CSM gas from the shocked stellar wind.
During the evolution of this structure, its outer part will collapse
due to the energy losses by emission and will appear as optical
nebula. On the other hand, its inner part is not subject to
significant energy losses due to its high temperature and low
density but its structure will depend on the efficiency of the
electron thermal conduction. In the case the thermal conduction 
is suppressed (e.g., due to presence of magnetic field), the hot bubble
will remain adiabatic and it will contain only the very hot gas
of the shocked stellar wind. 
Alternatively, the efficient thermal conduction will heat up part of 
the cold optical nebula which will expand inward (the contact
discontinuity will move towards the inner shock). As a result, the 
plasma temperature in the hot bubble will decrease and its mass will
increase. In this case the hot bubble will consist of two
parts both filled with hot gas but with different chemical
composition.

To derive the physical parameters of the hot bubble in \NGCE, we used
the 1D hydrodynamic code of A.V. Myasnikov that was developed for
modeling the physics of a standard WBB (for details see
\citealt{zhm_98}). This code allows modeling not only of adiabatic and 
radiative WBB but also of the case with efficient thermal conduction.
It correctly handles interacting gas flows with different chemical 
compositions.

We adopted the following basic set of parameters that determine
the physics of a WBB: 
(i) $\dot{M}_{WN} = 1.5\times10^{-5}$\dotM and $V_{WN} = 1600$\kms 
for the stellar wind of the central
star in \NGC (see \S~\ref{sec:thebubble});
(ii) $\dot{M}_{sw} = 1.5\times10^{-4}$\dotM and $V_{sw} = 10$\kms
for the slow wind that has been emitted in the previous evolutionary
stage of the central star.
The interaction of such winds results in a WBB with a radius of
$\sim 2.5$~pc at age of 30,000 years. 
We note that the size of the optical nebula \NGC is $12' \times 18'$
(\citealt{chu_83}; also see Fig.~\ref{fig:images}). Thus,
a spherically-symmetric (1D) bubble with a radius of $\sim
2.5$~pc has the same volume as a prolate ellipsoid with major and 
minor axes equal to those of the optical nebula for the adopted distance 
to this object ($d = 1.26$ kpc, \S~\ref{sec:thebubble}).
With this approximation, we calculate the distribution of emission measure 
of the hot bubble and the corresponding normalization parameter for
the model spectrum in XSPEC 
($ norm = 10^{-14} \int n_e n_H dV / 4 \pi d^2$). 
Then, the actual spectral
fit to the X-ray spectra can tell us if the theoretical emission
measure gives the correct value for the observed flux from \NGCE. 

\begin{figure}[ht]
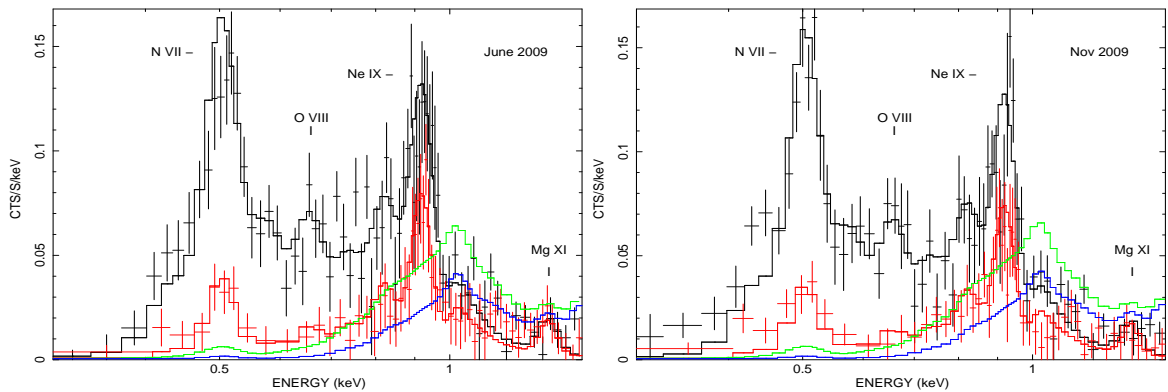

\centering\includegraphics[width=2.in,height=3.in,angle=-90]
{f6a.eps}
\centering\includegraphics[width=2.in,height=3.in,angle=-90]
{f6b.eps}
\caption{
\NGC background-subtracted spectra as in Fig.~\ref{fig:spectra}
overlaid with the best-fit model from the 1D hydrodynamic simulations
for conductive WBB with mass losses reduced by a factor of $\sim 4$.
In each panel, the XIS1 spectrum is in black and the XIS0$+$3 spectrum 
is in red (lower curve).
For comparison, the model spectra from a purely adiabatic WBB are shown 
in green (XIS1 spectrum) and blue (XIS0$+$3 spectrum). See text for
details.
}
\label{fig:spectra_wbb}
\end{figure}

We fitted the total spectra of \NGC with the theoretical X-ray
emission for the case of conductive WBB having the basic set of
physical parameters. In all model fits using results from our 
hydrodynamic simulations, the abundances of the shocked stellar wind
had values fixed to the typical WN abundances \citep{vdh_86} and
only abundances of the `evaporated' gas from the cold shell were
allowed to vary.
We note that using the basic set of parameters provided acceptable 
fit to the shape of the observed spectra but it grossly overestimated 
the observed flux (by a factor of  $\sim 50$). 
To reconcile this flux discrepancy, it was necessary to reduce the amount
of X-ray emitting plasma in the hot bubble. 
We ran a series of hydrodynamic simulations by simultaneously reducing
the mass-loss rates of the central star in \NGC and of the previously
emitted slow wind which results in no change of the global geometry of
the shock structure. In all these simulations, the wind velocities had
their values as in the basic set of physical parameters (see above).
The case with a reduction factor of $\sim 4$ for the mass-loss rates 
provided the correct observed flux and the derived
abundances  were a factor of (3 - 5) higher than the values in
Fig.~\ref{fig:abund}.
The corresponding spectral fit ($\chi^2/dof = 266/238$) to the 
{\it Suzaku} spectra of \NGC is shown in Fig.~\ref{fig:spectra_wbb}.

It is worth noting that although the required reduction for the mass-loss
rates might seem considerable, we find additional evidence that this 
could well be the case. If we assume that the thermal conduction is
not efficient and some other mechanism is responsible for the
relatively low plasma temperature deduced from the analysis of the
X-ray emission from \NGC (\S~\ref{sec:global}), the hot bubble in this
object will then be purely adiabatic. We ran a series of 1D
hydrodynamic models for this case starting with the same set of basic 
physical parameters as described above. 
The result was that the shocked stellar WN wind plasma had too 
high a temperature and the models provided strong X-ray flux at energies 
$> 1$~keV  considerably exceeding the one observed. 
For graphical 
clarity, we show in Fig.~\ref{fig:spectra_wbb} the theoretical spectrum 
of purely adiabatic WBB with  mass-loss rates reduced by a factor of 2. 
Note that the normalization (or the flux) of the theoretical spectrum 
scales with the mass-loss rate as $\propto \dot{M}^2$.

Therefore, it is conclusive that independently from which is the
heating mechanism for the X-ray emitting plasma in \NGCE,
we find indications that the mass-loss rate of the central star should
be smaller than the value of $\dot{M}_{WN} =
1.5\times10^{-5}$\dotM adopted here.
Direct comparisons of the model predictions and the X-ray observations
suggests a reduction factor of $\sim 3-4$.

%

\subsection{General Physical Picture}
\label{subsec:picture}
A general physical picture of the X-ray emission from a WBB must
provide a reasonable explanation about the origin of the X-ray
emitting gas and the mechanism that heats it up to the observed 
temperatures. Thus, we recall the basic properties of the X-rays 
from \NGCE.

\begin{figure}[ht]
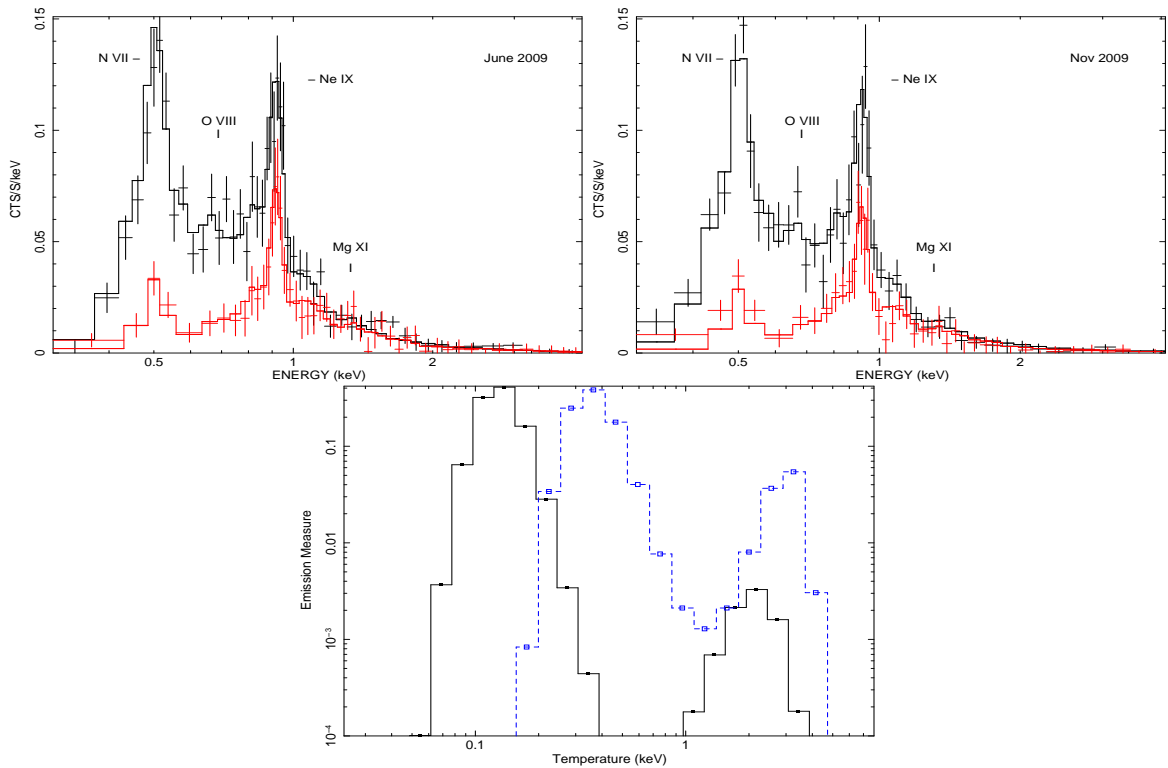

\centering\includegraphics[width=2.in,height=3.in,angle=-90]
{f5a.eps}
\centering\includegraphics[width=2.in,height=3.in,angle=-90]
{f5b.eps}
\centering\includegraphics[width=2.in,height=3.in,angle=-90]
{f5c.eps}
\caption{
The (0.3 - 4.0 keV) background-subtracted spectra of \NGC 
(re-binned to have a minimum of 200
counts per bin) overlaid with the best-fit model with a distribution
of CIE plasma.
In each panel, the XIS1 spectrum is in black and the XIS0$+$3 spectrum
is in red (lower curve).
The derived distribution of emission measure (normalized to have a
total sum of unity) is shown in the third
panel: that of the CIE plasma in black (solid line) and the one of the
NEI shocks in blue (dashed line). 
}
\label{fig:spectra_hot}
\end{figure}

%
The previous observations of the entire WBB with {\it ROSAT} and {\it
ASCA}
found that
the X-ray emission of \NGC comes from clump-like structures
(\citealt{wri_94}; \citealt{wri_05}). This is also the case as seen 
in the high signal-to-noise {\it Suzaku} data (Fig.~\ref{fig:images}).
Our analysis of the {\it Suzaku} spectra of \NGC showed that most of
the X-rays come from relatively cool plasma (kT $\leq 0.5$~keV) but
there is also some very hot plasma (kT $\geq 2$~keV;
Fig.~\ref{fig:spectra_hot}) that is responsible for the observed emission 
in the (1.5 - 4.0 keV) energy region. 

\begin{figure}[ht]
\centering\includegraphics[width=6.in,height=4.286in]{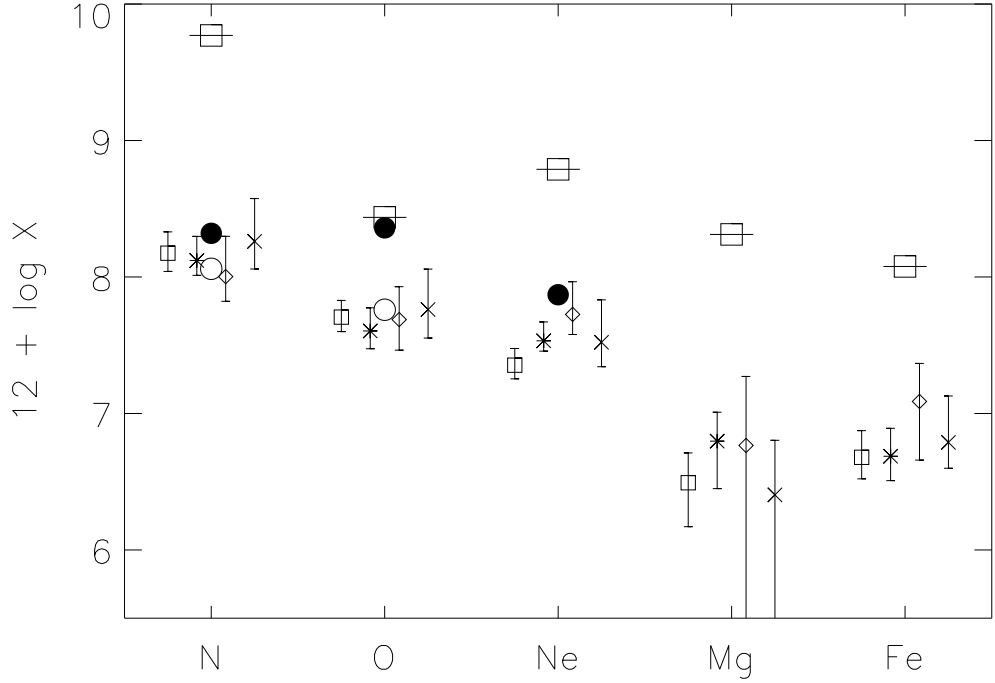}
\caption{
The abundance values of N, O, Ne, Mg and Fe derived from the global
model fits to the entire X-ray spectrum of \NGC.
The abundance values are with respect to the total 
number density of H$+$He and in units of $12 + \log X$ ($X$ is any of
N, O, Ne, Mg and Fe).
The symbols with error bars denote results from the fits with
{\it vpshock} and solar abundances - squares; {\it vpshock} and WN
abundances  - asterisks; distribution of CIE plasma - diamonds 
(the values from the 2T {\it vapec} model fits are not given since
they are within the plotted error bars); distribution of NEI shocks - 
crosses. Errors are $1\sigma$ values from the fits.
The abundances values derived from analysis of the optical data
by \citet{kwi_81} and \citet{moore_00} are shown with filled and empty
circles, respectively.
For reference, the {\it typical} abundances of WN stars \citep{vdh_86} 
are given by empty squares with horizontal bars (these bars have no
meaning and are used only for graphical clarity).
}
\label{fig:abund}
\end{figure}

The physical picture that emerges from the analysis of the
X-ray data of \NGC is then the following. 
Most of the soft X-rays come
from clumps distributed in the hot bubble of this object. 
As indicated by the similar chemical composition of the X-ray emitting
plasma and the optical shell (e.g., Fig.~\ref{fig:abund}), these clumps
likely originate from the cold  but ionized optical nebula.
Most probably they were formed
through various dynamic instabilities the optical nebula was subject
to during its evolution (e.g. see \citealt{gs_96a};
\citealt{gs_96b}; \citealt{stick_98};
and multiple shock waves may have contributed in the case of
conductive WBB, \citealt{zhm_98}). In addition, there is some very hot
plasma that can be associated with the shocked WN wind of the central
star. The details of this general physical picture depend on the
heating mechanism that operates in the hot bubble of \NGCE.

{\it Conductive WBB.}
If the electron thermal conduction is efficient, the energy of the
shocked stellar wind will be dissipated, part of the outer shell will
`evaporate' and this will form a smooth component of relatively hot
plasma filling in the inner part of the nebula. This case is well
represented by the hydrodynamic model of WBB with thermal conduction 
and this model gave an acceptable fit ($\chi^2/dof = 266/238$) 
to the observed spectra (see
\S~\ref{subsec:hydro} and Fig.~\ref{fig:spectra_wbb}). In this
picture, the dense ionized clumps that were formed via dynamic 
instabilities are engulfed by the hot bubble and heated up to X-ray 
temperatures thanks to the efficient thermal conduction. In fact, when 
we added an extra component (using model {\it vapec} in XSPEC) 
to the spectral fit based on the 
hydrodynamic simulations, the spectral model matched better the observed 
spectra ($\chi^2/dof = 232/238$) and the derived abundances had values 
consistent with those in Fig.~\ref{fig:abund}. The derived plasma 
temperature for this additional component (the clumps) was 
kT~$=0.11-0.12$~keV.
The main problem in the case of conductive WBB is that there cannot 
exist in the hot bubble plasma with temperatures as high as those 
derived from the fits to the {\it Suzaku} spectra in the (0.3 - 4.0 keV)
energy range (\S~\ref{subsec:hotplasma} and Fig.~\ref{fig:spectra_hot}).
As a result, practically no photons are produced in the (1.5 - 4.0
keV) energy band contrary to what is observed
(\S~\ref{subsec:hotplasma}).
We recall that {\it asymmetric} thermal conduction \citep{zhm_00}
might be a remedy to this problem if a global ordered magnetic field is
present in \NGCE. Thus, there might be a small part in the hot bubble
with a very hot plasma while the rest of it will be filled in with
much cooler plasma. Unfortunately, the number of detected photons in
the (1.5 - 4.0 keV) energy band did not allow us to study their
spatial distribution and much deeper observations are needed for that
purpose.

{\it Adiabatic WBB.} 
If the electron thermal conduction is suppressed
due to the presence of magnetic field, the hot bubble will be
adiabatic and the dense clumps that originate from the cold outer shell
and penetrate the bubble interior are heated up to X-ray
temperatures through some mechanism different from thermal conduction.
We explored quantitatively this case by considering a spectral model
consisting of two components: one representing the emission from the
adiabatic hot bubble (from our hydrodynamic simulations) and another one 
(using model {\it vapec} in XSPEC) representing the X-ray emission of 
the dense clumps. We obtained a good fit ($\chi^2/dof = 183/237$) to the 
{\it Suzaku} spectra of \NGC in the case of reduced by a factor of 3 
mass-loss rates. The temperature of the dense clumps was
kT~$=0.11-0.12$~keV and the derived abundances were consistent with
those in Fig.~\ref{fig:abund}. Then, what is the mechanism that heats 
up the dense clumps? Given the chemical composition of the clumps (or 
the cold nebula), shocks with velocity $\sim 300 - 350$\kms can provide the
required temperature (kT~$=0.11-0.12$~keV). But, simple considerations 
show that the rarefied plasma in the hot bubble does not have the 
necessary power to run such shocks into a gas with much higher density 
than the hot plasma itself (e.g., given the stellar wind velocity of 
1600\kms, the expected relative velocity between the hot plasma and the 
clumps is maximum of the order of the velocity of the shock). Could it 
be that the dense clumps mix {\it locally} with the hot plasma and since 
the electron thermal conduction is inefficient, the energy exchange
occurs directly between the heavy particles (ions) of the two
components? Consequently, the electrons in the mixture are heated as
well. Also, the relative velocity between the hot plasma and the clumps
will favor such a diffusion of the heavy particles.
We note that the adiabatic WBB does not have the high-temperature
caveat of the conductive WBB mentioned above.

But, how can we distinguish between the two cases, conductive or
adiabatic WBB? We think that there might be some observational
differences resulting from these two heating mechanisms.
The electron thermal conduction is a very efficient dissipative
mechanism, thus the heating of the cold gas of the clumps will occur 
almost 'instantaneously'. On the other hand, the ion-ion energy 
exchange (thanks to diffusion of the hot ions into the cold gas) is 
much slower process.
We recall that this different efficiency is due to the much
faster energy exchange between the light particles (the
characteristic time is proportional to the square root of the 
particle's mass; e.g., \citealt{spitzer_62}).
Thus, we may expect that there will exist clumps in the
hot bubble that have temperature intermediate between that of the 
optical outer shell and the X-ray clumps. These clumps will be source
of UV emission. Therefore, if spectral observations of \NGC detect 
strong lines in the 1,000-2,000~\AA ~spectral range 
(e.g., of `hot' ions like NV, OVI), this will indicate 
that the thermal conduction is considerably suppressed in this object.
Alternatively, if clumps with no strong UV emission are detected, this
will favor the thermal conduction as the heating mechanism in \NGCE.
Another observational possibility also exists, although it is not quite
feasible for the moment. \citet{no_09} reported a narrow radiative 
recombination continuum (RRC) in the X-ray spectrum of the planetary 
nebula BD$+60\degr3639$. They interpret this RRC emission 
feature as result when bare C VII ions penetrating from the hot bubble
into the cold optical nebula recombine with cool electrons. 
In general, the same can be expected when the hot ions diffuse into
the cold clumps as proposed above and RRC of various ions can be
expected. But, a crucial experiment for distinguishing between
detection from non-detection of these RRC features must wait for
future observations with the next generation of X-ray telescopes since
such an experiment requires high spectral resolution X-ray
observations which are not feasible for \NGC at the moment.

All this, as well as the suggested reduction for the mass-loss rate of 
the central WN star, emphasizes the importance to carry out some global 
analysis of the entire system \NGCE: central star, optical and X-ray
nebula (hot bubble). Namely, a self-consistent modeling of the
optical-UV emission of the central star (WR 136) and the optical
nebula and confronting it with observations can provide us with 
a valuable piece of information about the wind velocity and mass loss, 
temperature distribution in the stellar wind, its abundances as well
as the density and abundances of the optical nebula. This will put
much tighter constraints on the physical parameters determining the
physical conditions in the hot bubble (and its X-ray emission). In
turn, it  will result in a better constrained general physical picture 
of the whole system.
A suitable example of adopting such a global approach in the analysis
of the optical-UV emission of a complex system is given in
\citet{georgiev_08} for the case of the planetary nebula NGC~6543.

\subsection{Ionization History of the Hot Plasma}
\label{subsec:cie_nei}
The ionization history of the plasma in the hot bubble of a WBB is
quite complex. Consider the standard WBB model. In this case, the
thermal conduction is efficient and we have two parts in the hot
bubble filled correspondingly with the shocked stellar wind and the
evaporated gas from the cold optical nebula. During the evolution of
the WBB, the temperature in the hot bubble decreases with time as well
as does its density. Thus, each new parcel of gas, crossing the inner 
shock or evaporating from the cold nebula, will have its own ionization
history. This will depend on the time that parcel of gas entered the
hot bubble, that is on the physical parameters of the hot plasma and
their subsequent evolution until the moment of observations.
We note that it is likely (although it depends on the physical
parameters for each studied object) that the plasma on both sides 
in vicinity 
of the contact discontinuity will be almost in CIE while that close to
the inner shock or to the outer boundary in cold nebula will be in 
state of NEI.
In purely adiabatic WBB, the ionization history of the plasma is similar 
but then the hot bubble contains only the shocked stellar wind.
This picture becomes even more complicated in objects like \NGC where
the X-ray emission is not spatially uniform which indicates a presence
of clump like structures.

Unfortunately, it is not feasible to follow the ionization history of
the plasma in the hot bubble in the theoretical models or to deduce
valuable information about it from the X-ray observations. The latter
is particularly valid for the case of CCD X-ray spectra.
For that reason, our approach in this study was to consider two limiting
cases: plasma models with CIE and models that, although in a simplified
but well defined manner, take into account the NEI effects. 
Such an approach illustrates better the uncertainties in the derived
properties (e.g., temperature, abundances, emission measure) of the 
X-ray emitting plasma.

\section{Conclusions}
\label{sec:conclusions}
In this work, we presented an analysis of the {\it Suzaku} data on 
\NGCE: the wind-blown bubble around the Wolf-Rayet star WR 136. 
For the first time, X-ray spectra of the entire
nebula are obtained that have a very good quality allowing a detailed 
comparison between theory and observations. The basic results and 
conclusions are as follows.

\begin{enumerate}
\item
The X-ray spectra of the WBB \NGC are soft and most of the emission is 
in the (0.3 - 1.5 keV) energy range although some emission ($\sim
10$\% of the observed flux) is found at higher energies (1.5 - 4.0
keV). The spectral fits require a relatively cool plasma with 
kT~$\leq 0.5$~keV but much hotter plasma with temperature kT~$\geq
2.0$~keV is needed to match the observed hard X-ray emission.
We find no appreciable temperature variations within the hot bubble
of \NGCE.

\item
One of the important results from the spectral fits is that the
derived abundances (N, O, Ne) are very much the same as
those of the optical nebula. Also, they are considerably different from 
the WN abundances \citep{vdh_86} assumed `typical' for the wind of the
central star in the nebula.  This indicates a common origin of the
X-ray emitting gas and the outer cold shell, that is most of the X-ray 
plasma (likely concentrated in clumps as the X-ray images indicate) has 
flown into the hot bubble from the optical nebula.

\item
A direct comparison (in XSPEC) between 1D hydrodynamic models and the
X-ray spectra of \NGC suggests a reduced mass-loss rate ($\sim
4-5\times10^{-6}$\dotM) of the central star in order to provide the 
correct value of the observed flux. We note that this figure is
appreciably low for Wolf-Rayet stars and thus emphasizes the
need of a global modeling of the entire system: central star, optical
nebula and the hot bubble.

\item
The general physical picture that emerges from the current analysis is 
the following. Most of the X-rays come from clumps distributed in
the hot bubble of \NGCE. These clumps originate from the cold optical
nebula and they formed through various dynamic instabilities during
the evolution of the wind-blown bubble. The shocked WN wind of the
central star is likely the source of the very hot plasma (kT~$\geq
2$~keV) in this wind-blown bubble.
If the electron thermal conduction is efficient, this can naturally
explain the relatively low plasma temperature of most of the X-ray
emitting plasma. Alternatively, the hot bubble in \NGC will be
adiabatic and the cold clumps are heated up to X-ray temperatures
likely by energy exchange between the heavy particles (e.g., a local
diffusion of the rarefied hot plasma into the dense clumps).

\end{enumerate}



\begin{figure}[hp]
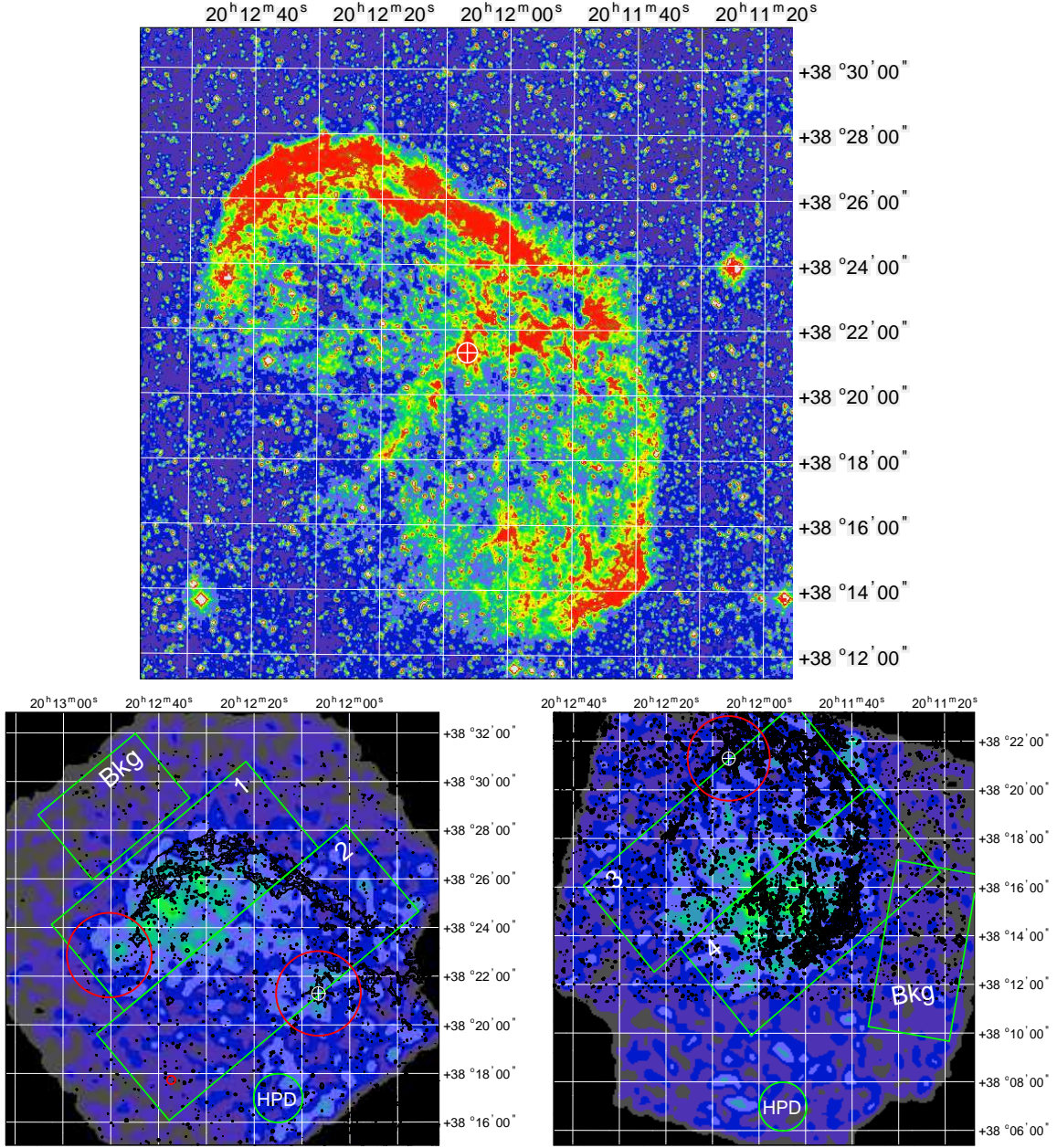

\centering\includegraphics[width=4.513in,height=4.in]{f1a.eps}
\centering\includegraphics[width=3.in,height=2.659in]{f1b.eps}
\centering\includegraphics[width=3.in,height=2.659in]{f1c.eps}
\caption{
\NGC images: optical (Palomar Observatory Sky Survey; upper panel); 
{\it Suzaku} XIS1 ($0.3-4.0$~keV) 
for the June
2009 (lower left panel) and November 2009 (lower right panel)
observations.
R.A. (J2000) and decl. (J2000) are on the horizontal and vertical axes,
respectively.
The boxes in green denote the extraction regions for the source 
(numbered from 1 through 4) and background spectra (labeled `Bkg').
The circles in red mark the regions around point sources that were
excluded from the spectral extraction. The circled plus sign gives the
position of the central star WR 136 in \NGC.
In lower panels, the circle in green (labeled `HPD') illustrates the
half-power diameter ($\sim 2$~arcmin) of the {\it Suzaku} telescopes.
The X-ray images are re-binned ($4\times4$ original pixels) and wavelet
smoothed for presentation. For clarity, overlaid are some contours
(in black) of the optical image.
Note that the lower panels show the {\it raw} X-ray images, thus,
their brightness scale is different due to the different non X-ray
background.
}
\label{fig:images}
\end{figure}
\begin{figure}[ht]
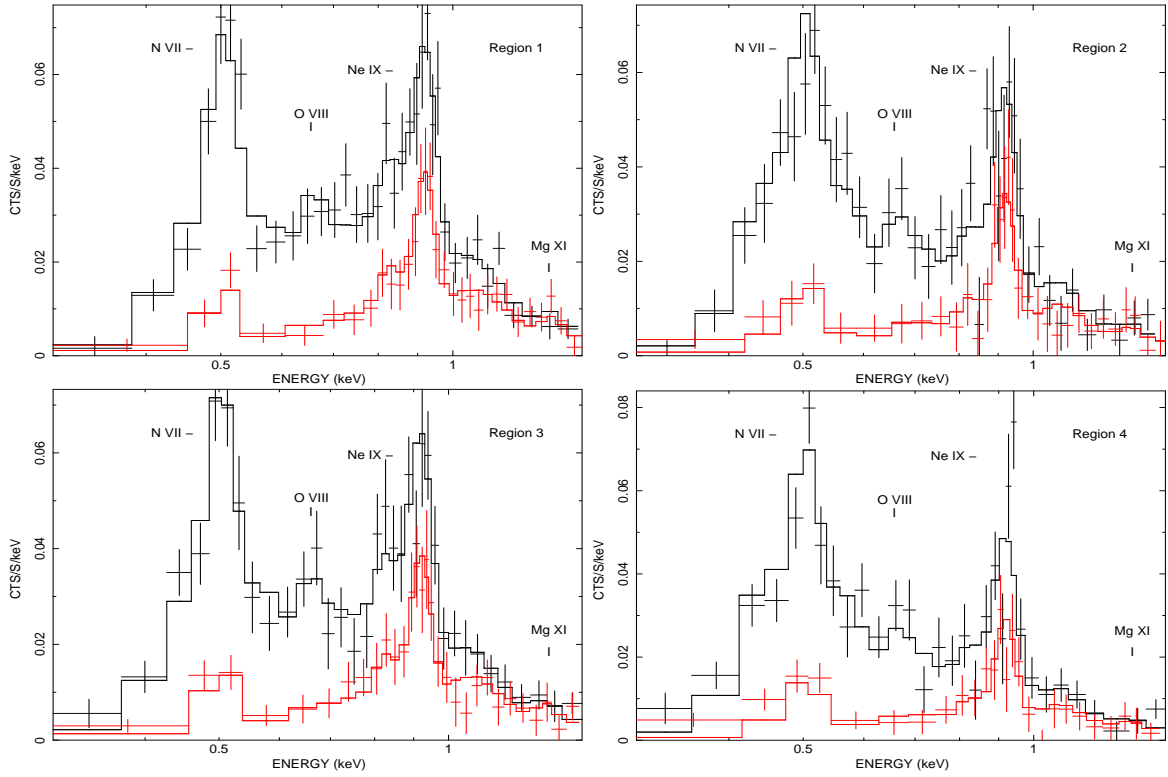

\centering\includegraphics[width=2.in,height=3.in,angle=-90]{f3a.eps}
\centering\includegraphics[width=2.in,height=3.in,angle=-90]{f3b.eps}
\centering\includegraphics[width=2.in,height=3.in,angle=-90]{f3c.eps}
\centering\includegraphics[width=2.in,height=3.in,angle=-90]{f3d.eps}
\caption{
\NGC background-subtracted spectra from the four extraction regions
in Fig.~\ref{fig:images} overlaid with the best-fit 1T shock model.
In each panel, the XIS1
spectrum is in black and the XIS0$+$3 spectrum is in red (lower
curve).
}
\label{fig:spectra_quart}
\end{figure}

\acknowledgments
This work was supported by NASA through the NASA Goddard award
NNX09AW39G to the University of Colorado at Boulder.
The optical image in Fig.~\ref{fig:images} is
based on photographic data of the National Geographic Society --
Palomar Observatory Sky Survey (NGS-POSS) obtained using the Oschin
Telescope on Palomar Mountain.  The NGS-POSS was funded by a grant from the
National Geographic Society to the California Institute of Technology.
The Digitized Sky Survey was produced at the Space Telescope Science 
Institute under US Government grant NAGW-2166.
The authors thank an anonymous referee for his/her comments 
and suggestions.



{\it Facilities:} \facility{{\it Suzaku} (XIS)}.

\clearpage



%

\clearpage

\clearpage

\clearpage

\clearpage

\clearpage

\clearpage

\begin{deluxetable}{lll}
\tablecaption{Global Spectral Model Results 
\label{tab:fits}}
\tablewidth{0pt}
\tablehead{
\colhead{Parameter} & \colhead{2T vapec}  & \colhead{vpshock} 
}
\startdata
$\chi^2$/dof  & 172/232 & 177/235  \\ 
N$_{H}$ (10$^{21}$ cm$^{-2}$)  & 
          1.92$^{+0.42}_{-0.44}$ & 2.07$^{+0.47}_{-0.46}$ \\ 
kT$_1$ (keV) & 0.15$^{+0.01}_{-0.01}$ & 0.48$^{+0.22}_{-0.11}$ \\ 
kT$_2$ (keV) & 2.03$^{+1.31}_{-0.52}$ & ......  \\ 
EM$_1$ ($10^{55}$~cm$^{-3}$) &  80.8$^{+25.6}_{-16.6}$ &  
                                6.3$^{+3.3}_{-1.8}$ \\
EM$_2$ ($10^{55}$~cm$^{-3}$) &  1.0$^{+0.17}_{-0.14}$  & 
                                ...... \\
Abundances  & Varied & Varied  \\ 
$\tau$ ($10^{10}$ cm$^{-3}$ s)  &   &   3.98$^{+2.85}_{-1.52}$ \\ 
F$_X$ ($10^{-12}$ ergs cm$^{-2}$ s$^{-1}$)  & 
           2.05 (18.4) & 1.92 (17.4) \\ 
\enddata
\tablecomments{
Results from  simultaneous fits to the XIS1 and XIS0$+$3 
spectra of \NGC from the {\it Suzaku} observations in June and Nov 2009.
Tabulated quantities are the neutral hydrogen absorption column
density (N$_{H}$), plasma temperature (kT), 
emission measure ($\mbox{EM} = \int n_e n_H dV $), 
shock ionization
time-scale ($\tau = n_e t$), the X-ray flux (F$_X$) in the 
0.3 - 1.5 keV range followed in parentheses by the unabsorbed value.
Only abundances of N, O, Ne, Mg and Fe were varied
in the fits (see Fig.~\ref{fig:abund}).
Errors are the $1\sigma$ values from the fits.
}

\end{deluxetable}




\end{document}